# PlumX As a Potential Tool to Assess the Macroscopic Multidimensional Impact of Books

*Daniel Torres-Salinas[1,2], Christian Gumpenberger[3] and Juan Gorraiz[3]\**

[1] *Universidad de Navarra, Navarra, Spain,* [2] *Universidad de Granada (EC3metrics spin off y Medialab UGR), Granada, Spain,*
[3] *Bibliometrics and Publication Strategies, Vienna University Library, University of Vienna, Vienna, Austria*

The main purpose of this macro-study is to shed light on the broad impact of books. For this purpose, the impact of a very large collection of books (more than 200,000) has been analyzed by using PlumX, an analytical tool providing a great number of different metrics provided by various tools. Furthermore, the study also describes the changes in the values of the most significant measures and indicators over time. The results show that the usage counts in comparison to the other metrics are quantitatively predominant. Catalog holdings and reviews represent a book's most characteristic measures deriving from its increased level of impact in relation to prior results. Our results also corroborate the long half-life of books within the scope of all metrics, excluding views and social media. Despite some disadvantages, PlumX has proved to be a very helpful and promising tool for assessing the broad impact of books, especially because of how easy it is to enter the ISBN directly as well as its algorithm to aggregate all the data generated by the different ISBN variations.

## INTRODUCTION

Initially, monographs were the primary channels used for scholarly communication. However, the evolution from "little to big science" resulted in an increased level of competition and required research findings to be published faster than before. Due to these new developments, books finally lost their principal role as a publication channel, particularly in the fields of the natural and exact sciences. Even if the future of scholarly monographs might be uncertain (Williams et al., 2009; Maynard and O'Brien, 2010; Watkinson et al., 2016), they are still indispensable academic outputs (Nederhof, 2006; Huang and Chang, 2008) in the arts, humanities, and some social sciences. This is corroborated by the 2014 UK Research Excellence Framework statistics (REF, 2014). According to these figures, book submissions are still common in the humanities (55%), in the arts (33%), and in the social sciences (22%), compared to a vanishing 0.5% in science, engineering, and medicine (Kousha et al., 2016).

Despite their continuing importance to the disciplines mentioned beforehand, books are still looked down upon within the scope of evaluation exercises or practices. The absence of books in key bibliometric databases has resulted in an unfortunate depreciation of monographs perceived to be scientific products of minor value. Due to the pressure exerted by national evaluation policies and international conventions, many researchers are shifting or have switched from books to journal articles as their preferred dissemination channel, which has been experienced in the UK (Research Information Network, 2009).

Furthermore, university rankings, including the ones primarily based on bibliometric data like the Leiden Ranking, tend to ignore books. Unfortunately, this also holds true for the disciplines previously mentioned with books still playing a crucial role. Only a few bibliometric analyses with evaluative purposes exist today that consider books as publication channels in these fields and all of them corroborate their indisputable importance and significant contribution to citation analyses (e.g., Kousha and Thelwall, 2009; Gorraiz et al., 2016).

Therefore, the impact assessment of monographs is a big challenge and a hot topic in the scientometric field. Citation analysis can be an acceptable proxy for measuring publication impact, but only for a limited subset of the scientific community, namely the "publish or perish" group and only regarding the impact reflected by documented scholarly communication. However, it is a common knowledge that many disciplines address much broader audiences within and even beyond the scholarly community (societal impact). Thus, monographs can generate educational or public interest rather than, or in addition to, research impact (Kousha and Thelwall, 2015a,b). Moreover, they potentially aim to culturally enrich a non-academic audience (Small, 2013).

Certainly, impact needs to be defined on a much broader scale and new metrics have the potential to paint a more complete picture concerning the impact generated by monographs. These new metrics aim to complement (or even replace) traditional citation-based metrics. On one hand, usage metrics are drawn from the usage data collected (e.g., views and downloads), which are now available due to all the licensed electronic journals, books, and other content (Gorraiz et al., 2014a; Glänzel and Gorraiz, 2015). On the other hand, altmetrics reflect the interest or attention that research outputs have generated in the social web (Priem and Hemminger, 2010; Priem et al., 2010; Robinson-Garcia et al., 2014). The increasing adoption of Web 2.0 practices and the ongoing development and launch of innovative tools are promising for the application of alternative evaluation approaches, which will finally take books into consideration as well.

The launch of the book citation index in 2011 resulted in the higher and quicker accessibility of citation data for large collections of books. It also triggered several citation studies to be conducted on books' citation patterns, characteristics, and peculiarities (e.g. Kousha et al., 2011; Leydesdorff and Felt, 2012; Torres-Salinas et al., 2012, 2013, 2014; Gorraiz et al., 2013). Additionally, in order to solve the shortcomings of citations, other proxies for assessing monographs have been suggested in literature based on:

- library holdings such as the number of catalog entries per book title in WorldCat® (Torres-Salinas and Moed, 2009), library bindings (Linmans, 2010), and even introducing an indicator of perceived cultural benefit (White et al., 2009)
- document delivery requests (Gorraiz and Schlögl, 2006)
- library loans (Cabezas-Clavijo et al., 2013)
- publishers' prestige (Torres-Salinas et al., 2012, 2013; Giménez-Toledo et al., 2013)
- book reviews (Nicolaisen, 2002; Zuccala and Van Leeuwen, 2011; Gorraiz et al., 2014b; Bornmann, 2015; Kousha and Thelwall, 2015a,b; Zuccala et al., 2015; Kousha et al., 2016; Thelwall and Kousha, 2016; Zhou et al., 2016)
- altmetrics (Zhou and Zhang, 2013).

Currently, the following three major tools collect and aggregate altmetrics data: ImpactStory, Altmetric.com, and PlumX. Altmetric.com and PlumX focus on institutional customers (e.g., publishers, libraries, and universities) by gathering and providing data on large scale, and ImpactStory rather targets individual researchers who wish to include altmetrics information in their CV (Peters et al., 2016). For this study, we use the fee-based PlumX altmetrics dashboard for the following reasons:

- it gathers and offers publication-level metrics for so-called artifacts, which also include, apart from monographs or books, articles, audios, videos, book chapters, patents, or clinical trials;
- it allows ISBNs to be directly entered as well as many other identifiers (IDs) ranging from user IDs—such as ORCID, or other more specific ones (e.g., YouTube and SlideShare)—to a large number of publication IDs, such as DOIs, PubMed-IDs, URLs, and patent numbers; and
- due to its integration into the EBSCO Platform, PlumX can provide statistics on the usage of e-books and other artifacts (e.g., views to or downloads of HTML pages or PDFs).

The provider of PlumX™ is Plum Analytics, a 100% subsidiary of EBSCO Information Services since 2014. However, during the course of writing this article, Elsevier took over PlumX from EBSCO. Early in 2012, Plum Analytics was founded by Andrea Michalek and Mike Buschman with the vision of bringing modern ways of measuring research impact on individuals and organizations that use and analyze research. Furthermore, PlumX provides its own analytics consisting of the following three different standard reports: artifacts by publication year, artifacts overview, and sunburst (**Figure 1**).

Unfortunately, there are no studies to date that take some or all of these proxies simultaneously into account, thereby allowing for a multidimensional perspective. This is mainly due to the fact that obtaining data are cumbersome and time-consuming. However, comparison of results and identifying relevant indicators and their interaction would be highly desirable. Therefore, this study is a first explorative attempt toward a multidimensional approach at a macro level. In this pilot analysis, we use a case study based on a large collection of books provided by the University of Vienna.

## RESEARCH QUESTIONS

The main purpose of this study is to shed light on the impact of books. On one hand, the possibilities of PlumX to analyze the impact of books will be closely examined; on the other hand, the impact of books will then be analyzed by means of a large number of different metrics provided by different tools and to thereby achieve the intended multidimensional character. In this first macro-study, we focus on following research questions:

(1) *Methodological and technical questions*: how can we calculate book metrics with PlumX, and which is the most effective way? Can significant differences in the results be observed according to the used book ID during data retrieval? Is PlumX a valid platform for book impact assessment, and what main technical limitations arise?

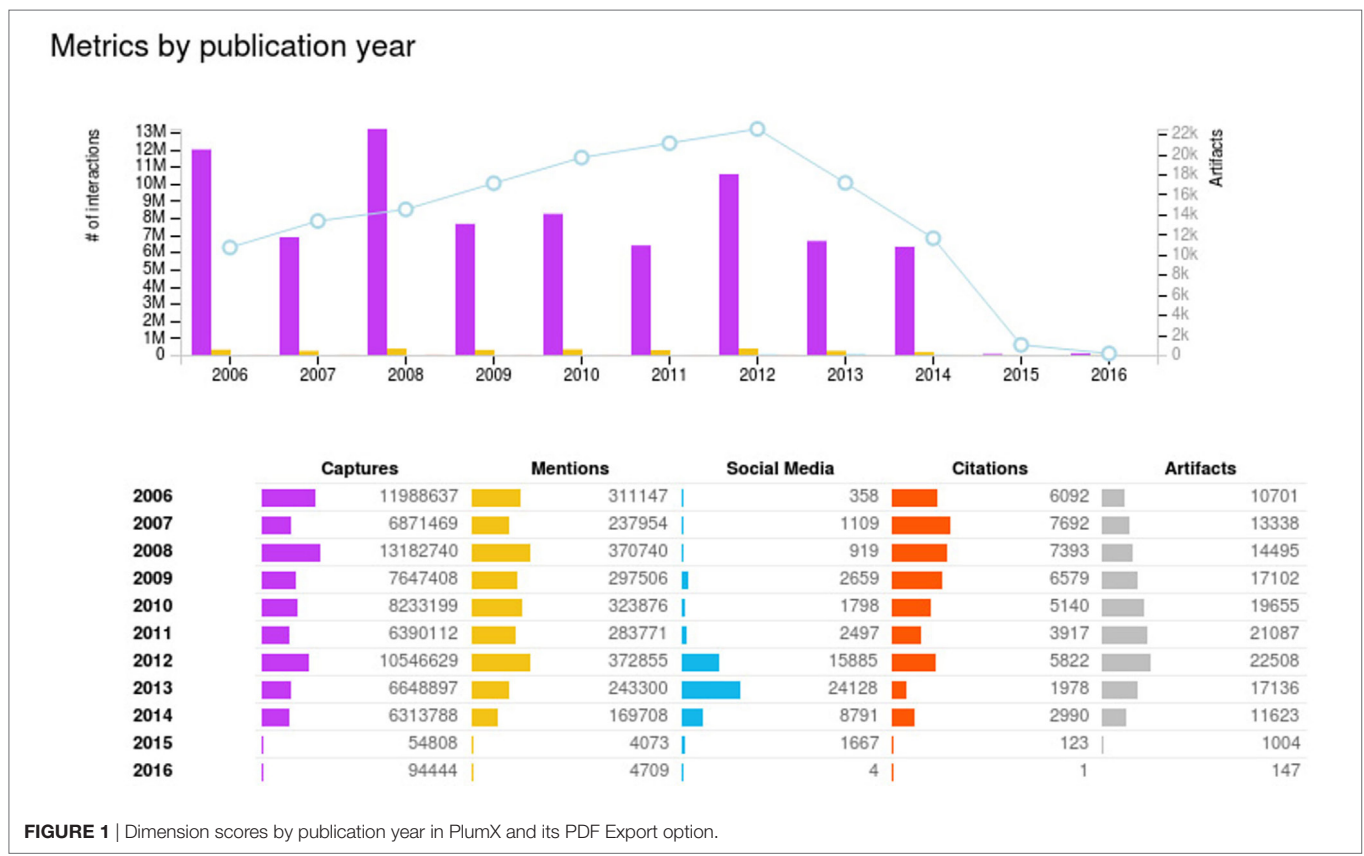

**FIGURE 1** | Dimension scores by publication year in PlumX and its PDF Export option.

(2) *Indicators analysis questions*: related to metrics we can address the following issues: which measures are provided by PlumX for books? Which are most relevant and significant? What tools do they originate from? Which influences can be observed in the different measures collected during evolution over time concerning the years of publication years of the books analyzed?

The article is organized as follows: In the Section "Methodology and Technical Questions," we explain the data origin and the process of data collection. In the Section "Results," we describe the measures generated by each information source and analyze their statistical characteristics. Finally, in the Section "Discussion and Conclusion," we discuss the relevance of the obtained results, the appropriateness of our approach versus the main limitations of the study, as well as other technical issues that could potentially affect future bibliometric analyses assessing the broad impact of books.

## METHODOLOGY AND TECHNICAL QUESTIONS

In this section, we describe the dataset used as input data, data processing by PlumX—including detailed information of the resulting dataset (output)—the measures and indicators retrieved in this tool and the analyses that we have performed in order to address the research questions of this study.

## Dataset: EBSCO Book Collection— University of Vienna

For this analysis, we used the complete book collection provided by EBSCO and licensed by the Vienna University Library as our defined dataset. It contains 268,061 books. According to the universal character of the University of Vienna, the EBSCO academic e-book collection contains titles from a broad range of subject and topic areas (from science to performing arts), different book categories (such as fiction, non-fiction, and reference), as well as a large number of publishers (e.g., De Gruyter, Nova Science Publishers, Routledge, and Oxford University Press). Furthermore, the publication years of the book collection range from 1,800 until now. According to language, English was predominant with 230,050 titles (almost 86%), followed by German (8.4%), French (1.9%), and Spanish (1.8%).

Overall, the dataset included books in more than 70 different languages. Moreover, 114,290 titles were single volumes of serials (42.6%) and 153,771 were monographs (57.4%). Finally, we have included Complementary Material (available at: http://hdl.handle.net/10481/45114) in order to enrich the analysis and provide the reader with further information.

### Artifacts and Indicators

In order to gather the data in PlumX, a plain text file containing all the ISBNs of the sample has been introduced. After entering our input data, PlumX processes the books and provides a new dataset including all the resulting "artifacts," as items are named in PlumX in the output. This resulting dataset can be exported to Excel in CSV format, and this is the final dataset that we use for the analysis. The dataset structure differed considerably from the input dataset; for our purpose, an analysis on the macro level, the outgoing data can be used taking into account different considerations. Since the input data are not entirely reflected in the obtained output data, the resulting "artifacts" have been re-classified into different types, such as articles, papers, references, books, book chapters, and datasets, and enriched with further bibliographical attributes (e.g., ISBN, titles, publication years, and DOI).

The resulting dataset also includes the scores of all measures according to their origin. **Table 1** lists all measures and indicators according to the sources that have been retrieved from. The measures are categorized into the following five separate dimensions: usage, captures, mentions, social media, and citations (**Figure 1**; **Table 1**). The data retrieved originate from EBSCO sources as well as from other external platforms. In total, 23 different measures were available, including traditional metrics, such as citations, as well as usage data and altmetrics. Thus, a much multidimensional picture of the broad impact of the books is painted. EBSCO is the prevalent source of provided usage data.

This categorization may be subject to criticism, but one big advantage of PlumX is that the results are differentiated in the resulting dataset for each measure and its origin and can be aggregated according to the user criterion.[1] Furthermore, PlumX provides its own analytics consisting of the following three different standard reports: artifacts by publication year, artifacts overview, and sunburst. In our study and in order to perform statistical analyses, we used the complete resulting dataset, performed our own analysis, and opted for our own representation of the results.

## Analyses

Preliminary analyses were conducted in order to shed light on any effects observed that were caused in the output dataset while using print or electronic ISBNs as input data. Both analyses were performed simultaneously at the end of November 2016 and after the same period of data consolidation. **Table 3** shows the results of the artifacts categorization automatically performed in PlumX in both cases. Concerning document types, 260,856 titles (99.1% of the data) were correctly identified as books by PlumX. Less than 1% comprises non-identified artifacts (0.7%), references (0.06%), articles (0.06%), or book chapters (0.05%). **Table 2** also shows that the percentage of the total retrieved artifacts is higher when using the ISBN of the print version, but the correct artifact identification as books is slightly higher when using ISBN related to the electronic version.

The PlumX results were very similar in both ISBN. The number of scores retrieved was only slightly higher by using the ISBN of the print version, while only the number of clicks was higher when using the ISBN of the electronic version. From the total number of books introduced by using only the ISBN of the print version being identified and processed by PlumX as books (260,854, see **Table 2**), 23.83% were identified by the ISBN of the

---

[1]In order to better understand how PlumX processes the data, a detailed input versus output analysis was performed for all the books from the year 2011 ($n = 22,378$). The ISBNs of the print version of all these books were used as input data. From the total amount of books introduced, 88.34% (19,768) were identified and processed by PlumX and only 28 (less than 0.001%) were identified as other document types than books. After a cumbersome work of manual disambiguation, 19,305 of these 19,768 (around 98%) were identified according to the ISBN of the print version (32%) or the ISBN of the electronic version (68%). The rest of the books (2%) were almost correctly assigned to the title book but according to the

ISBN of other formats, editions, or catalog holdings. Furthermore, 16,590 of these 19,305 books were assigned to the correct publication year of the book (85.94%), 1,733 books were assigned to wrong publication years or publication years of other editions (8.98%), and 982 books were assigned to any publication year (5.09%).

**TABLE 1** | Measures gathered in PlumX classified according to their dimension and sources.

| Dimension usage | Dimension mentions | Dimension social media |
|---|---|---|
| **Abstract views**→Dspace | **Reviews**→Amazon | **Tweets**→Twitter |
| • *Source type*: repository | • *Source type*: electronic bookseller | • *Source type*: microblogging network |
| **Downloads**→Dspace | **Reviews**→Goodreads; Social | **Shares, likes and comments**→Facebook |
| • *Source type*: repository | • *Source type*: cataloging website | • *Source type*: social platform |
| **Sample downloads**→EBSCO | **News mentions**→News | **Score**→Reddit |
| • *Source type*: electronic provider | • *Source type*: online reference | • *Source type*: social platform |
| **Abstract views**→EBSCO | **Links**→Wikipedia | **Likes**→Google+ |
| • *Source type*: electronic provider | • *Source type*: online reference | • *Source type*: social platform |
| **Data views**→EBSCO | | |
| • *Source type*: electronic provider | **Dimension captures** | **Dimension citations** |
| **PDF views**→EBSCO | **Export saves**→EBSCO | **Citation counts**→CrossRef |
| • *Source type*: electronic provider | • *Source type*: electronic provider | • *Source type*: database |
| **HTML views**→EBSCO | **Readers**→Mendeley | **Citation counts**→Scopus |
| • *Source type*: electronic provider | • *Source type*: reference manager | • *Source type*: database |
| **Link-outs**→EBSCO | **Readers**→Goodreads | **Citation counts**→PubMed |
| • *Source type*: electronic provider | • *Source type*: social platform | • *Source type*: database |
| **Holdings**→WorldCats | | **Citation counts**→Others |
| • *Source type*: libray catalog | | • *Source type*: database |

print version and 51.38% were identified by the ISBN of the electronic version (even if these were not used as data input). Only 0.44% was identified according to both ISBN variants and the rest of the books, 25.15%, were identified according to other versions or variants of the ISBN not even available in the original dataset. There was also a discrepancy between the input and output data, concerning years of publication as it is shown and compared with the temporal distribution of the input data in **Table 1**. And 9.7% of the retrieved data by PlumX did not contain any information about the year of publication. **Figure 2** provides a general view of the methodology and data gathering process.

The metrics analysis was performed for the large collection of books used as input data and the resulting PlumX dataset by using the ISBN of the print versions of the books. In order to correctly address our second research question, we considered only the document type "book" in the output data (**Figure 2**), in order to avoid data originated by other document types and not the book itself. In order to study the temporal evolution of the indicators, analyses were performed for the following five different specific time periods: before 2000, 2000–2003, 2004–2007, 2008–2011, and 2012–2015. To avoid incorrectly assigning the years of publication, we used the parts of the resulting dataset (output)

**TABLE 2** | Artifact categorization in PlumX according to ISBN of the print or electronic version.

| PlumX output | ISBN (print) | ISBN (electronic) |
|---|---|---|
| Artifact | Total artifacts | Total artifacts |
| Book | 260,856 | 261,225 |
| Book chapter | 125 | 123 |
| Article | 145 | 49 |
| Conference paper | 10 | 7 |
| Paper | 6 | 0 |
| Reference | 171 | 155 |
| Report | 11 | 9 |
| Textual work | 2 | 2 |
| Research artifact | 1,884 | 460 |
| **Total** | **263,210** | **262,030** |
| All artifacts% input | 98.19% | 97.75% |
| Books% input | 97.31% | 97.45% |

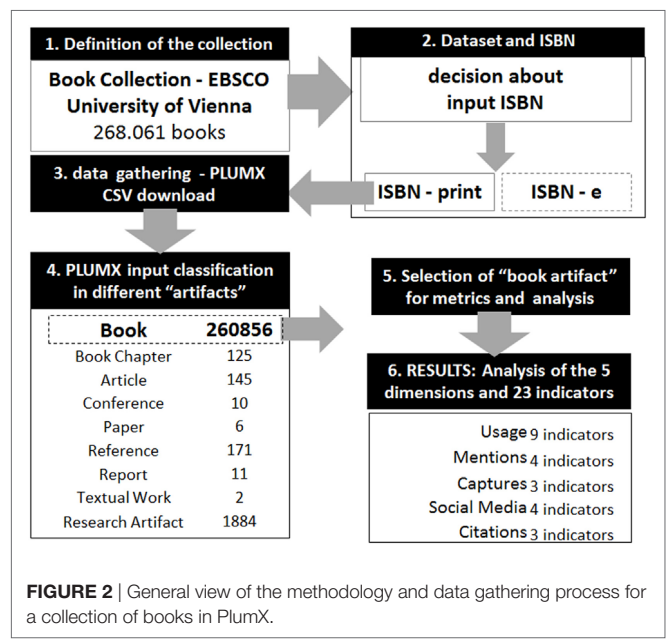

**FIGURE 2** | General view of the methodology and data gathering process for a collection of books in PlumX.

**TABLE 3** | Statistical overview for the dimensions and main measures retrieved *via* PlumX according to the ISBN of the print version of the book.

|  |  | Books with data available | Book with data available (%) | Sum | Mean | Max value | SD |
|---|---|---|---|---|---|---|---|
| **(A) Descriptive statistics for the five dimensions** | | | | | | | |
| Total captures | | 238,040 | 91.25 | 167,747,227 | 643.06 | 3,355,710 | 22,471.76 |
| Total citations | | 11,448 | 4.39 | 250,551 | 0.96 | 4,260 | 22.01 |
| Total social media | | 2,183 | 0.84 | 60,917 | 0.23 | 12,420 | 32.18 |
| Total mentions | | 123,305 | 47.27 | 4,778,855 | 18.32 | 55,044 | 395.76 |
| Total usage | | 258,977 | 99.28 | 322,303,716 | 1,235.56 | 1,423,424 | 4,167.24 |
| **(B) Descriptive statistics for main measures in PlumX—coverage >1%** | | | | | | | |
| Usage | Holdings:WorldCat | 255,137 | 97.81 | 184,766,294 | 708.31 | 35,276 | 619.06 |
| Usage | Abstract views:EBSCO | 248,135 | 95.12 | 94,549,206 | 362.04 | 648,337 | 2,085.53 |
| Captures | Exports saves:EBSCO | 213,299 | 81.77 | 7,241,548 | 27.76 | 93,507 | 262.02 |
| Usage | HTML views:EBSCO | 197,657 | 75.77 | 24,113,722 | 92.44 | 768,653 | 2,041.39 |
| Usage | PDF views:EBSCO | 178,101 | 68.28 | 17,291,795 | 66.29 | 31,853 | 281.75 |
| Captures | Readers:Goodreads | 140,088 | 53.70 | 160,074,381 | 613.65 | 3,355,704 | 22,469.90 |
| Usage | Link-outs:EBSCO | 139,216 | 53.37 | 1,325,721 | 5.08 | 4,175 | 25.81 |
| Mentions | Reviews:Amazon | 97,182 | 37.26 | 1,551,583 | 5.95 | 18,149 | 70.43 |
| Captures | Readers:Mendeley | 64,842 | 24.86 | 431,296 | 1.65 | 944 | 8.72 |
| Mentions | Reviews:Goodreads | 49,893 | 19.13 | 3,105,890 | 11.91 | 51,422 | 359.70 |
| Mentions | Links:Wikipedia | 43,216 | 16.57 | 119,805 | 0.46 | 931 | 4.19 |
| Usage | Sample downloads:EBSCO | 39,470 | 15.13 | 74,109 | 0.28 | 146 | 1.13 |
| Citations | Citation indexes:CrossRef | 11,080 | 4.25 | 232,557 | 0.89 | 4,260 | 21.00 |
| Usage | ePub downloads:EBSCO | 2,861 | 1.10 | 47,320 | 0.18 | 1,378 | 6.28 |

*For more information and complete data, check Table S1 in Supplementary Material.*

according to the publication years assigned by PlumX. The results are listed, compared, and discussed in the next sections.

## RESULTS

### Dimensions and Indicators in PlumX

The results are summarized in **Table 3** and **Figure 3** for the most relevant measures collected from each source and grouped according to PlumX dimensions including a short statistical analysis in order to provide enough information about the distribution of the scores.

The upper right section of **Figure 3** represents the measures with higher data availability and volume. They corroborate the quantitative predominance of usage counts in comparison to the other metrics. Around 98% of the books were cataloged in WorldCat. A total of 95% of the abstracts and 68% of the PDF files were downloaded *via* EBSCO. It should be noted that the total values for each dimension or category group were only calculated in order to give a quick overview of the percentage of books with available data.

The highest percentage of books with the data on captures is reported by EBSCO (around 82%)—as expected considering the structure of the sample—followed by Goodreads (more than half). Only approximately one-fourth of the books were "captured" in Mendeley. The highest number of readers is, however, reported by Goodreads with a mean value of more than 600 for the complete sample and a mean value of more than 1000 if we only considered books with at least one score. The number of readers in Mendeley is insignificant in comparison. The mean value of export saves in EBSCO is around 28. The distribution is very skew according to the values of the maxima and SD in comparison with other measures like citations.

The number of cited books was extremely low in all the citation indexes available in PlumX: just 4.25% in CrossRef and only 0.12% in Scopus. The mean number of citations of the total sample was less than 0.9 per book; the mean number of citations of the books with available data was 21 and 54 for Crossref and Scopus, respectively. Note that Scopus citation data were retrieved only for 316 books. The presence of books in social media is even lower. Twitter and Facebook reported the highest percentages but less than 1%. The distribution is also extremely shifted by a few outliers. The percentage of books with mentions is lower than the one for captures but significantly higher than the one for cited books or for social media. A total of 37% of the books have at least a review in Amazon and 19% in Goodreads. However, Goodreads shows the highest mean value of review scores per book.

Furthermore, around 17% of the books were mentioned and linked in Wikipedia. The mean number of catalog holdings was almost double as the number of abstract views and 10 times higher than the number of downloads but comparable to the median number of readers in Goodreads. One citation is related to 28 captures, 66 PDF views, or 92 HTML views *via* EBSCO and around 10 and 6 reviews in Goodreads and Amazon, respectively.

Spearman correlations were then calculated for all these parameters with a significant number of data available (Table S2 in Supplementary Material). These are Exports-Saves:EBSCO, Readers:Mendeley, Readers:Goodreads, Scopus, CrossRef, Tweets:Twitter, Facebook, Reviews:Amazon, Reviews:Goodreads, Links:Wikipedia, Abstract Views:EBSCO, Holdings:WorldCat, PDF Views:EBSCO, and HTML Views:EBSCO.

No calculations were made for total scores of any of the categorical groups because they were only used as an aid in order to quickly estimate the percentage of the data availability for each category. However, due to the heterogeneous mix of data and aspects, mathematical sums are not of any relevance in calculating correlations. The results show the very low and almost inexistent correlation between almost all parameters, except for some measures that originated from the same tools, such as EBSCO—export saves with abstract and HTML views—and Goodreads—number of readers with number of reviews.

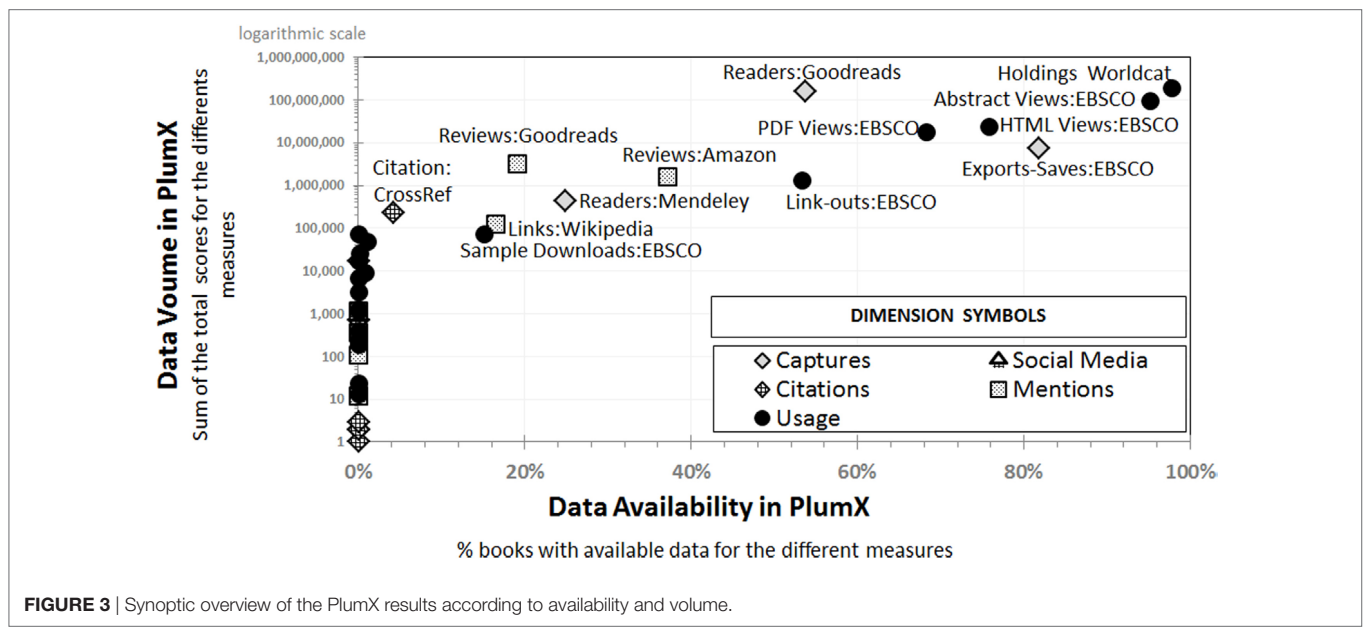

**FIGURE 3** | Synoptic overview of the PlumX results according to availability and volume.

## Comparison of Different Intervals Publication Years

A detailed statistical overview of the results obtained for all the books published before 2000, 2000–2003, 2004–2007, 2008–2011, and 2012–2015 (according to the publication years of the output set) is given in tables $x$, $x + 1$, $x + 2$, $x + 3$, and $x + 4$ (Tables S3–S7 in Supplementary Material). **Figure 4** provides an overview of the percentage of data available for the most significant scores during the five selected time periods.

The availability of usage data is the highest and remains almost constant over the five periods. All the other percentage of data availability for downloads, captures, reviews, mentions, and citations shows a maximum for the interval 2000–2003. These results corroborate the long half-life of books in all metrics, excluding views and social media. For social media, the percentage of data availability is strongly increasing in agreement with the advance and increasing popularity of this new metric in the last years (see the Section "Results" for the period 2012–2015), but the percentage of books with social media scores is still very low in this most recent period: 1% in Twitter and 1.34% in Facebook.

**Figure 5** also shows the trend line of the mean values of the available data for the most significant parameters. Very similar results were obtained when considering the mean values of the total sample and not the mean of the available data (mean instead "available as supplementary material" for Tables 7–11). It shows that the mean value of the social media scores (yellow lines) increases as strongly as the one of the citation counts (red lines). The mean value of the usage scores (green lines) also decreases along the time but not so sharply as the citation counts. The mean values of captures (blue lines) and mentions (orange lines) fluctuate but seem to reach a maximum in the case of books older than 10 years (e.g., with publication years between 2004 and 2007). Again, it should be noted that the total values for each dimension or category group were only calculated in order to give a quick and general overview. Therefore, the results for the most relevant main measures retrieved *via* PlumX are provided in **Table 4**.

## DISCUSSION AND CONCLUSION

### Methodology and Technical Limitations

The results of this large-scale study help to identify the different shapes of the broad impact of books. PlumX allows us to calculate a large number of metrics quickly. This tool retrieves

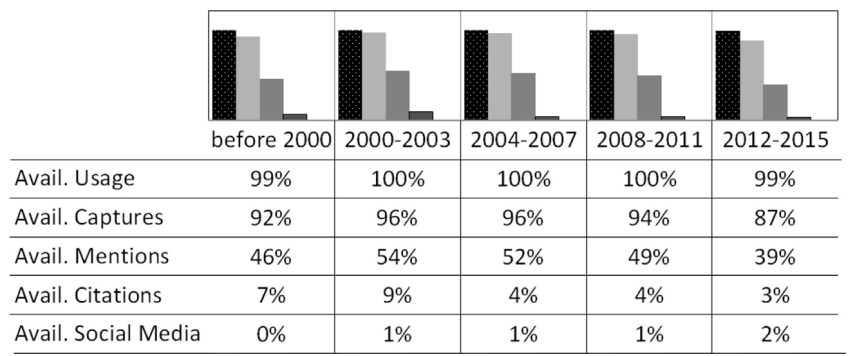

|                    | before 2000 | 2000-2003 | 2004-2007 | 2008-2011 | 2012-2015 |
|--------------------|-------------|-----------|-----------|-----------|-----------|
| Avail. Usage       | 99%         | 100%      | 100%      | 100%      | 99%       |
| Avail. Captures    | 92%         | 96%       | 96%       | 94%       | 87%       |
| Avail. Mentions    | 46%         | 54%       | 52%       | 49%       | 39%       |
| Avail. Citations   | 7%          | 9%        | 4%        | 4%        | 3%        |
| Avail. Social Media| 0%          | 1%        | 1%        | 1%        | 2%        |

**FIGURE 4** | Percentage of availability and mean value for each dimension over the time. Book with data available (%) forms the five dimensions in PlumX. For more information and complete data, check Tables S3–S7 in Supplementary Material.

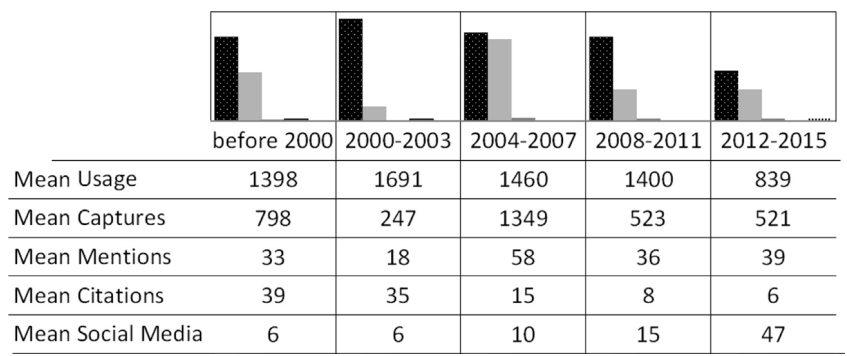

|                   | before 2000 | 2000-2003 | 2004-2007 | 2008-2011 | 2012-2015 |
|-------------------|-------------|-----------|-----------|-----------|-----------|
| Mean Usage        | 1398        | 1691      | 1460      | 1400      | 839       |
| Mean Captures     | 798         | 247       | 1349      | 523       | 521       |
| Mean Mentions     | 33          | 18        | 58        | 36        | 39        |
| Mean Citations    | 39          | 35        | 15        | 8         | 6         |
| Mean Social Media | 6           | 6         | 10        | 15        | 47        |

**FIGURE 5** | Percentage of availability and mean value for each dimension over the time. Mean value for the five dimensions in PlumX. For more information and complete data, check Tables S3–S7 in Supplementary Material.

**TABLE 4** | Percentage of availability and mean value for main measures and tool over the time.

| | PlumX indicator | Time period | | | | |
|---|---|---|---|---|---|---|
| | | Before 2000 | 2000–2003 | 2004–2007 | 2008–2011 | 2012–2015 |
| **(A) Book with data available (%) form main indicators in PlumX** | | | | | | |
| Usage | Holdings:WorldCat | 98.87% | 99.14% | 99.04% | 98.56% | 95.73% |
| Usage | Abstract views:EBSCO | 92.34% | 96.45% | 96.28% | 96.56% | 95.87% |
| Captures | Exports saves:EBSCO | 82.30% | 89.43% | 88.22% | 86.82% | 76.23% |
| Usage | HTML views:EBSCO | 73.13% | 83.68% | 77.25% | 79.96% | 72.31% |
| Usage | PDF views:EBSCO | 74.28% | 84.26% | 80.39% | 76.92% | 53.76% |
| Captures | Readers:Goodreads | 56.01% | 61.74% | 57.52% | 55.95% | 46.80% |
| Usage | Link-outs:EBSCO | 46.08% | 58.46% | 55.02% | 57.38% | 54.85% |
| Mentions | Reviews:Amazon | 33.77% | 42.40% | 40.73% | 38.78% | 31.61% |
| Captures | Readers:Mendeley | 25.25% | 33.53% | 30.31% | 27.41% | 22.94% |
| Mentions | Reviews:Goodreads | 14.99% | 18.25% | 20.04% | 21.74% | 17.68% |
| Mentions | Links:Wikipedia | 20.70% | 24.17% | 21.01% | 16.23% | 9.83% |
| Usage | Sample downloads:EBSCO | 7.48% | 9.16% | 10.41% | 14.84% | 25.80% |
| Citations | Citation indexes:CrossRef | 6.52% | 9.21% | 4.03% | 3.63% | 3.01% |
| Usage | ePub downloads:EBSCO | 0.27% | 0.37% | 0.36% | 0.86% | 2.07% |
| **(B) Mean value for main indicators in PlumX** | | | | | | |
| Usage | Holdings:WorldCat | 923.43 | 1,004.12 | 820.52 | 750.35 | 451.51 |
| Captures | Readers:Goodreads | 711.66 | 195.97 | 1,254.75 | 527.03 | 434.31 |
| Usage | Abstract views:EBSCO | 321.18 | 455.61 | 440.51 | 434.49 | 267.49 |
| Usage | HTML views:EBSCO | 69.82 | 109 | 98.85 | 133.64 | 73.44 |
| Usage | PDF views:EBSCO | 72.01 | 111.48 | 89.45 | 75.12 | 28 |
| Captures | Exports-saves:EBSCO | 22.53 | 37.79 | 34.06 | 36.07 | 18.86 |
| Mentions | Reviews:Goodreads | 11 | 4.86 | 20.91 | 11.78 | 9.6 |
| Mentions | Reviews:Amazon | 3.58 | 4.35 | 8.56 | 5.77 | 5.4 |
| Usage | Link-outs:EBSCO | 2.98 | 4.85 | 5.2 | 5.1 | 6.23 |
| Captures | Readers:Mendeley | 1.77 | 2.91 | 2.28 | 1.63 | 1.16 |
| Citations | Citation indexes:CrossRef | 2.52 | 3.13 | 0.54 | 0.25 | 0.11 |
| Mentions | Links:Wikipedia | 0.64 | 0.76 | 0.62 | 0.39 | 0.21 |
| Usage | Sample downloads:EBSCO | 0.12 | 0.15 | 0.17 | 0.17 | 0.57 |
| Usage | ePub downloads:EBSCO | 0.05 | 0.03 | 0.06 | 0.1 | 0.44 |

*For more information and complete data check Tables S3–S7 in Supplementary Material.*

26 measures or indicators, a very broad abundance compared to other platforms or tools, for example, Altmetric.com that allows downloading just 18.

It is also interesting to note the assignment parameters for grouping metrics into five non-hierarchical dimensions or categories against a synthetic indicator like the Altmetric Score (Gumpenberger et al., 2016). However, this categorization into five dimensions is of course not completely accurate and free of criticism due to their heterogeneity, including different measures or indicators from different tools and potential overlapping. Therefore, simple score additions at this level should be avoided and only used to provide a quick idea of the data availability in each one of them.

Furthermore, according to PlumX, data processing will be performed in 3 h; however, this might depend on the size of the data sample. In our study, the data were introduced at the beginning of October 2016 and controlled during the next 2 months in order to guarantee that all of the data are fully processed. No considerable differences were observed after 2 weeks. For samples with less than 50,000 items, the processing was normally completed within 3 h. Another advantage of this tool is allowing a very quick and simple entry of large datasets in comparison with other products, such as Web of Science and Scopus, thereby limiting the number of items to be downloaded.

One shortcoming of PlumX is that the form of the input data is not conserved and precisely depicted in the output data. After processing of the data, the tool provides other bibliographical information (ISBN, titles, publication years, DOI, etc.), which are not always in perfect agreement with the bibliographic attributes of the input. Thus, not all the output records were conserving the same ISBN used for the entry data and a very arduous manual disambiguation was required in order to get an exact match between the input and output data.

Our results hint toward the hypothesis that PlumX would create its own data index of all books that have already been processed by this tool at least once, automatically enhancing their bibliographical data, and, for example, in the case of books including different variations of the ISBN already used as identification in one of the previous searches (Figure S1 in Supplementary Material).

In our particular study, concerning books, another additional problem arises because of the different uses of the ISBN. The Standard Book Number is assigned to each edition and variation of a book.[2] The OCLC (also named OCN) control number, a unique number associated with a record in WorldCat, is also not a satisfactory solution for this problem because it also depends on the format of the book rather than the title or content. By the way, OCNs is not as universal as ISBNs.[3]

---

[2]For example, an e-book, a paperback and a hardcover edition of the same book would each have a different ISBN.

Actually, the ISBN policy is oriented according to the publishers' and booksellers' requirements and not according to other user groups, such as librarians and scientometricians, being more interested in tracing the whole data of a book title than in identifying each format. Therefore, a generic permanent ID, which encompasses all the e-book and other formats for a title, is still lacking. Such a parent or global ID would be very suitable to meet the requirements of the Digital Era and the assessment of the impact of such an important channel of scholarly communication as books are in many disciplines.

In the meantime, only the ability of the specific tools to identify and collect other related ISBNs can help. To this respect, PlumX seems to have developed a very helpful and promising algorithm in order to aggregate all the data generated by the different ISBN of each version for each book title, according to the entry data delivered in prior searches and indexed in their own database.[4]

A correct identification of the artifacts is always very cumbersome as it has been reported even for journal articles under the use of the DOI, PubMed ID, and other document IDs repeatedly.

## Indicators Analysis

The results reinforce the crucial importance of the usage metrics in order to assess the impact of books. One of the most relevant aspects of this study is the differentiation of the five dimensions and measures in two antagonistic groups: the ones with good coverage and higher values (usage and captures) in comparison to the other ones with contrary behavior (social media and citations).

A comparison with previous results for journal articles shows that the percentage of books' downloads is even higher than the one reported for journal articles being around 90% (Gorraiz et al., 2014a). However, the percentage of cited books remains much lower than the one reported for contributions to journals, fluctuating between 40 and 60%. Furthermore, according to this prior study for journal articles—which considered a publication window of 10 years—the download mean frequency is at least double as higher for journal articles (between 50 and 140 downloads) than for books (approximately 26 and 29 for monographs).

Another big advantage of PlumX is the consideration of indicators or measures that are very closely related to the nature of books, such as catalog holdings. Around 98% of the books were cataloged in WorldCat. The mean number of catalog holdings was almost double as the number of abstract views and 10 times higher than the number of downloads and only comparable to the median number of readers in Goodreads. These results corroborate that catalog holdings are also very specific and usual data type for books. Undoubtedly, they shed light on the potential usage of the books and are very useful for the impact assessment of books as already suggested in prior studies (Torres-Salinas and Moed, 2009; Gorraiz et al., 2011).

Reviews are also one of the most characteristic footprints of a book, and the data originating from tools such as Amazon and Goodreads are certainly of crucial importance as has already been reported in prior studies (Zuccala and Van Leeuwen, 2011; Gorraiz et al., 2014b; Zuccala et al., 2015). More than half of the books were "captured" in Goodreads while only one-fourth in Mendeley. Amazon includes reviews from more than one-third of the books considered in this analysis, while Goodreads is less than 20%. Furthermore, only a median correlation between the number of reviews in Amazon and Goodreads was reported. These results have been corroborated by previous studies performed by Kousha et al. (2016), Thelwall and Kousha (2016). They reported the existence of significant but low or moderate correlations between citations in WoS and numbers of reviews based on a set of 2,739 academic monographs from 2008 and a set of 1,305 best-selling books in 15 Amazon.com academic subject categories (Kousha and Thelwall, 2016). Our results corroborate their suggestion that online book reviews tend to reflect the wider popularity of a book rather than its academic impact, although there are substantial disciplinary differences.

Concerning intrinsically altmetrics, the presence of books in social media was even lower. Twitter and Facebook reported the highest percentages but less than 1%. The distribution is also extremely shifted by few outliers. However, our study shows that the presence of books in social media is strongly increasing in the last years in agreement with the increasing popularity and advancement of these tools in recent years.

The almost inexistent correlation between all the measures and indicators emphasizes the multidimensionality of all the compiled information. The low correlation of usage data with citations and social media data is in agreement with de Winter's analysis of around 30,000 *PLoS ONE* articles revealing that the number of tweets was weakly associated with the number of citations and that the number of tweets was predictive of other social media activities (e.g., Mendeley and Facebook) but not of the number of article views on PubMed Central (de Winter, 2015). Interesting is also the almost non-existent correlation between EBSCO exports and the number of readers neither in Mendeley nor in Goodreads. This hints at quite different user communities.

Our results also corroborate the long half-life of books in all metrics, excluding views and social media. The percentage of data availability and the mean scores values for downloads, captures, reviews, mentions, and citations show a peak for books published almost 13–16 years previously.

## Final Remarks

According to these results, PlumX has proven to be a very useful tool in order to trace the impact of footprints (imprints) of a book in different areas according to the new metrics at the macro level. Our analysis strengthens the philosophy of the tool PlumX providing a cornucopia of measures grouped in different groups but not providing a simple and composite indicator. In doing so, the multidimensional aspect is better addressed, even if it is far from trivial in dealing with such an amalgam of different types of information retrieved from a plethora of data sources (Gumpenberger et al., 2016).

---

[3]For example, in our study, only 91% of the books included in our sample have an OCN.

[4]However, for the impact assessment of an individual book, it will be recommended to use all the ISBNs according to the different formats available in order to guarantee a complete data retrieving. Arduous manual work will be required in order to exclude duplicates. Some tools, such as PlumX, enable tracing the origin of each score in order to exclude potential duplication of score counts, but this function is not available or even possible for all metrics (e.g., usage metrics, downloads, and views) and in all tools generating the data.

The great benefit of this tool entails retrieving the scores resulting from each tool separately and enabling different methods of aggregation or interpretation and different statistical analyses of the data. The quite different aspects of scholarly communication retrieved from all the tools available in PlumX need to be separately analyzed and discussed, in order to preserve the multidimensionality and complexity of the different metrics and indicators.

According to our results—very low correlation even in the same dimension or category—different tools might provide only partial views and they should be considered rather as complementary sources in order to reach a higher completeness of data. Differentiation according to language, book types, and disciplines will provide a better understanding of the impact of books and factors to be considered for its correct assessment. Studies on these topics are ongoing.

Last but not least, further research might attempt to clarify the stability and reproducibility of altmetrics data, in order to get a thorough and transparent documentation of their temporal evolution and to trace and understand potential score changes.

It should also be stressed that the kind of book impact analysis required by university rankings or similar evaluation exercises focuses on a particular set of books and is different from the macroscopic impact analysis of the whole book collections as described in this article.

## AUTHOR CONTRIBUTIONS

JG is responsible for idea, concept, data retrieval, and data interpretation. CG is responsible for literature research, context, formulation, and conclusions. DT-S is responsible for data management (retrieval, cleaning, etc.), representation of the data (figures and tables), interpretation, and conclusions.

## ACKNOWLEDGMENTS

The authors thank Stephan Buettgen (EBSCO) for granted trial access to PlumX and Benedikt Blahous for his help processing the data.

## SUPPLEMENTARY MATERIAL

The Supplementary Material for this article can be found online at http://journal.frontiersin.org/article/10.3389/frma.2017.00005/full#supplementary-material.

## REFERENCES

Bornmann, L. (2015). Usefulness of altmetrics for measuring the broader impact of research: a case study using data from PLOS and F1000Prime. *Aslib J. Info. Manage.* 67, 305–319. doi:10.1108/AJIM-09-2014-0115

Cabezas-Clavijo, A., Robinson-García, N., Torres-Salinas, D., Jiménez-Contreras, E., Mikulka, T., Gumpenberger, C., et al. (2013). *Most Borrowed is not Most Cited? Library Loan Statistics as a Proxy for Monograph Selection in Citation Indexes.* Available at: http://arxiv.org/abs/1305.1488

de Winter, J. C. F. (2015). The relationship between tweets, citations, and article views for PLOS ONE articles. *Scientometrics* 102, 1773–1779. doi:10.1007/s11192-014-1445-x

Giménez-Toledo, E., Tejada-Artigas, C., and Mañana-Rodríguez, J. (2013). Evaluation of scientific books' publishers in social sciences and humanities: results of a survey. *Res. Eval.* 22, 64–77. doi:10.1093/reseval/rvs036

Glänzel, W., and Gorraiz, J. (2015). Usage metrics versus altmetrics: confusing terminology? *Scientometrics* 102, 2161–2164. doi:10.1007/s11192-014-1472-7

Gorraiz, J., Gumpenberger, C., and Glade, T. (2016). On the bibliometric coordinates of four different research fields in geography. *Scientometrics* 107, 873–897. doi:10.1007/s11192-016-1864-y

Gorraiz, J., Gumpenberger, C., and Schloegl, C. (2014a). Usage versus citation behaviours in four subject areas. *Scientometrics* 101, 1077–1095. doi:10.1007/s11192-014-1271-1

Gorraiz, J., Gumpenberger, C., and Purnell, P. J. (2014b). The power of book reviews: a simple and transparent enhancement approach for book citation indexes. *Scientometrics* 98, 841–852. doi:10.1007/s11192-013-1176-4

Gorraiz, J., Gumpenberger, C., and Wieland, M. (2011). Galton 2011 revisited: a bibliometric journey in the footprints of a universal genius. *Scientometrics* 88, 627–652. doi:10.1007/s11192-011-0393-y

Gorraiz, J., Purnell, P. J., and Glänzel, W. (2013). Opportunities for and limitations of the book citation index. *J. Am. Soc. Inf. Sci. Technol.* 64, 1388–1398. doi:10.1002/asi.22875

Gorraiz, J., and Schlögl, C. (2006). Document delivery as a source for bibliometric analyses: the case of Subito.2. *J. Inf. Sci.* 32, 223–237. doi:10.1177/0165551506064410

Gumpenberger, C., Glänzel, W., and Gorraiz, J. (2016). The ecstasy and the agony of the altmetric score. *Scientometrics* 108, 977–982. doi:10.1007/s11192-016-1991-5

Huang, M., and Chang, Y. (2008). Characteristics of research output in social sciences and humanities: from a research evaluation perspective. *J. Am. Soc. Inf. Sci. Technol.* 59, 1819–1828. doi:10.1002/asi.20885

Kousha, K., and Thelwall, M. (2009). Google book search: citation analysis for social science and the humanities. *J. Am. Soc. Inf. Sci. Technol.* 60, 1537–1549. doi:10.1002/asi.21085

Kousha, K., and Thelwall, M. (2015a). "Alternative metrics for book impact assessment: can choice reviews be a useful source?," in *Proceedings of the 15th International Conference on Scientometrics and Informetrics*, Vol. 5, 9–70. Available at: https://www.semanticscholar.org/paper/Alternative-Metrics-for-Book-Impact-Assessment-Can-Kousha-Thelwall/c69a38d1ac5bafd750a3f54411452e8ac6d5f79d

Kousha, K., and Thelwall, M. (2015b). Web indicators for research evaluation. Part 3: books and non-standard outputs. *El profesional de la información* 24, 724–736. doi:10.3145/epi.2015.nov.04

Kousha, K., and Thelwall, M. (2016). Can Amazon.com reviews help to assess the wider impacts of books? *J. Assoc. Inf. Sci. Technol.* 67, 566–581. doi:10.1002/asi.23404

Kousha, K., Thelwall, M., and Abdoli, M. (2017). Goodreads reviews to assess the wider impacts of books. *J. Assoc. Inf. Sci. Technol.* doi:10.1002/asi.23805

Kousha, K., Thelwall, M., and Rezaie, S. (2011). Assessing the citation impact of books: the role of Google Books, Google Scholar, and Scopus. *J. Am. Soc. Inf. Sci. Technol.* 62, 2147–2164. doi:10.1002/asi.21608

Leydesdorff, L., and Felt, U. (2012). Edited volumes, monographs and book chapters in the book citation index (BKCI) and science citation index (SCI, SoSCI, A&HCI). *J. Scientometric Res.* 1, 28–34. doi:10.5530/jscires.2012.1.7

Linmans, A. J. M. (2010). Why with bibliometrics the humanities does not need to be the weakest link. Indicators for research evaluation based on citations, library bindings and productivity measures. *Scientometrics* 83, 337–354. doi:10.1007/s11192-009-0088-9

Maynard, S., and O'Brien, A. (2010). Scholarly output: print and digital – in teaching and research. *J. Doc.* 66, 384–408. doi:10.1108/00220411011038467

Nederhof, A. (2006). Bibliometric monitoring of research performance in the social sciences and the humanities: a review. *Scientometrics* 66, 81–100. doi:10.1007/s11192-006-0007-2

Nicolaisen, J. (2002). The scholarliness of published peer reviews: a bibliometric study of book reviews in selected social science fields. *Res. Eval.* 11, 129–140. doi:10.3152/147154402781776808

Peters, I., Kraker, P., Lex, E., Gumpenberger, C., and Gorraiz, J. (2016). Research data explored: an extended analysis of citations and altmetrics. *Scientometrics* 107, 723–744. doi:10.1007/s11192-016-1887-4

Priem, J., and Hemminger, B. H. (2010). Scientometrics 2.0: new metrics of scholarly impact on the social web. *First Monday*. doi:10.5210/fm.v15i7.2874

Priem, J., Taraborelli, D., Groth, P., and Neylon, C. (2010). *Altmetrics: A Manifesto*. Available at: http://altmetrics.org/manifesto/

REF. (2014). *Results and Submissions in Research Excellence Framework*. Available at: http://results.ref.ac.uk/

Research Information Network. (2009).

Robinson-Garcia, N., Torres-Salinas, D., Zahedi, Z., and Costas, R. (2014). New data, new possibilities: exploring the insides of Altmetric.com. *El Profesional de la Información* 23, 359–366. doi:10.3145/epi.2014.jul.03

Small, H. (2013). *The Value of the Humanities*. Oxford, UK: Oxford University Press. ISBN: 9780199683864.

Thelwall, M., and Kousha, K. (2016). Goodreads: a social network site for book readers. *J. Assoc. Inf. Sci. Technol.* Available at: http://www.scit.wlv.ac.uk/~cm1993/papers/GoodReadsASocialNetworkSiteForBookReaders_preprint.pdf

Torres-Salinas, D., and Moed, H. F. (2009). Library catalog analysis as a tool in studies of social sciences and humanities: an exploratory study on published book titles in economics. *J. Inf.* 3, 9–26. doi:10.1016/j.joi.2008.10.002

Torres-Salinas, D., Robinson-García, N., Fernandez-Valdivia, J., and García, J. A. (2014). Analyzing the citation characteristics of books: edited books, book series and publisher types in the book citation index. *Scientometrics* 98, 2113–2127. doi:10.1007/s11192-013-1168-4

Torres-Salinas, D., Robinson-García, N., Jiménez-Contreras, E., and Delgado López-Cózar, E. (2012). Towards a 'book publishers citation reports'. First approach using the 'book citation index'. *Revista Española de Documentación Científica* 35, 615–620. doi:10.3989/redc.2012.4.1010

Torres-Salinas, D., Rodriguez-Sánchez, R., Robinson-García, N., Fdez-Valdivia, J., and García, J. A. (2013). Mapping citation patterns of book chapters using the book citation index. *J. Inf.* 7, 412–424. doi:10.1016/j.joi.2013.01.004

Watkinson, A., Nicholas, D., Thornley, C., Herman, E., Jamali, H. R., Volentine, R., et al. (2016). Changes in the digital scholarly environment and issues of trust: an exploratory, qualitative analysis. *Inf. Process. Manag.* 52, 446–458. doi:10.1016/j.ipm.2015.10.002

White, H. D., Boell, S. K., Yu, H., Davis, M., Wilson, C. S., and Cole, F. T. (2009). Libcitations: a measure for comparative assessment of book publications in the humanities and social sciences. *J. Am. Soc. Inf. Sci. Technol.* 60, 1083–1096. doi:10.1002/asi.21045

Williams, P., Stevenson, L., Nicholas, D., Watkinson, A., and Rowlands, I. (2009). The role and future of the monograph in arts and humanities research. *Aslib Proc.* 61, 67–82. doi:10.1108/00012530910932294

Zhou, Q., and Zhang, C. (2013). Relationship between scores and tags for Chinese books in the case of Douban book. *Chin. J. Libr. Inf. Sci.* 6, 40–54.

Zhou, Q., Zhang, C., Zhao, S. X., and Chen, B. (2016). Measuring book impact based on the multi-granularity online review mining. *Scientometrics* 107, 1435–1455. doi:10.1007/s11192-016-1930-5

Zuccala, A., and Van Leeuwen, T. (2011). Book reviews in humanities research evaluations. *J. Am. Soc. Inf. Sci. Technol.* 62, 1979–1991. doi:10.1002/asi.21588

Zuccala, A. A., Verleysen, F. T., Cornacchia, R., and Engels, T. C. (2015). Altmetrics for the humanities: comparing goodreads reader ratings with citations to history books. *Aslib J. Inf. Manag.* 67, 320–336. doi:10.1108/AJIM-11-2014-0152

**Conflict of Interest Statement:** The authors declare that the research was conducted in the absence of any commercial or financial relationships that could be construed as a potential conflict of interest.